\PassOptionsToPackage{sort}{natbib}  
\documentclass[reprint,prl,aps]{revtex4-1}
\usepackage{hyperref}

\usepackage{mathabx}

\usepackage{amsmath}
\usepackage{amsfonts}
\usepackage{bm}
\usepackage{graphicx}
\begin{document}
\title{Decoherence of black hole superpositions}
\author{Andrew Arrasmith}
\affiliation{Theory Division, LANL, Los Alamos, New Mexico 87545, USA}
\affiliation{Department of Physics, University of California Davis, Davis, California, 95616, USA}
\author{Andreas Albrecht}
\affiliation{Center for Quantum Mathematics and Physics and Department of Physics, University of California Davis, Davis, California, 95616, USA}
\author{Wojciech H. Zurek}
\affiliation{Theory Division, LANL, Los Alamos, New Mexico 87545, USA}
\begin{abstract}
We consider the decoherence of a ``black hole Schr{\"o}dinger cat'' -- a non-local superposition of a Schwarzschild black hole in two distinct locations -- due to the Hawking radiation 
it inevitably emits. An environment interacting with a system acquires information about its state, e.g. about its location. The resulting decoherence is thought to be responsible for the emergence of the classical realm of our Universe out of the quantum substrate. However, this view of the emergence of the classical is sometimes dismissed as a consequence of insufficient isolation and, hence, as non-fundamental (i.e., for practical purposes only). 
A black hole can never be isolated from its own Hawking radiation environment that carries information about its location. The resulting decoherence rate turns out to be given by a surprisingly simple equation that, remarkably (and in contrast to known cases of decoherence), does not involve Planck's constant $\hbar$.
\end{abstract}
\date{\today}

\maketitle

\section*{\label{sec:intro}Introduction}
Despite nearly a century of effort, the unification of quantum mechanics with general relativity is still a work in progress. 
Perhaps the most promising breakthrough in investigating the relation between quantum theory and general relativity came when Hawking \cite{Hawking} (following heuristic arguments of Bekenstein \cite{Bekenstein}) used quantum field theory to show that Schwarzschild black holes radiate as if they were at a temperature $T_H$ given by
\begin{eqnarray}
\label{eq:T_H}
k_B T_H=\frac{\hbar c^3}{8 \pi G M} =\frac{\hbar c}{4 \pi R_s},
\end{eqnarray}
where $M$ is the mass of the black hole and $R_s$ is the Schwarzschild radius.

Hawking radiation defies the classical expectation that nothing can be emitted from a black hole.  It was initially hoped that this result would pave the way to quantization of gravity. However, Hawking radiation has instead deepened the mystery by implicating entropy (and, hence, information) in questions involving quantum theory and gravity (e.g. the black hole information paradox). 

The origin of classicality in other settings has been, in the meantime, clarified by the theory of decoherence\cite{Joos_book,Zurek_review,Schlosshauer}. As in the black hole information paradox, information plays a key role: Decoherence is caused by the information flowing from the system into its environment and the resulting formation of records of its selected observables in that environment \cite{Zurek_Pointer,Zurek_Einselection,Zurek_Darwin}. It is now widely (though not universally) accepted that the effectively classical behavior of macroscopic systems in our quantum Universe is a consequence of decoherence. Even weak interactions can result in such leakage of information, and, consequently, in decoherence. 
However, information has been traditionally viewed as inconsequential in fundamental classical physics. Therefore, the decoherence-based view of the emergence of “the classical” in our quantum Universe has been regarded by some \cite{Bell}
as not fundamental.

Unlike most other cases that have been investigated, a black hole is a system that cannot be isolated: it creates its own environment -- Hawking radiation. Therefore, its decoherence is not just a practical matter, but fundamental. The decoherence of black holes by Hawking radiation thus provides a glimpse into a place where quantum mechanics and general relativity meet while a quantum-to-classical transition is happening. Below we obtain and discuss the rate of decoherence of a Schwarzschild black hole in both thermal equilibrium  with a radiation bath as well as in a vacuum and discuss the implications of our results.

\section*{\label{sec:thermal}Decoherence in a thermal bath}
\begin{figure}
 \begin{center}
   \includegraphics[width=.4\textwidth]{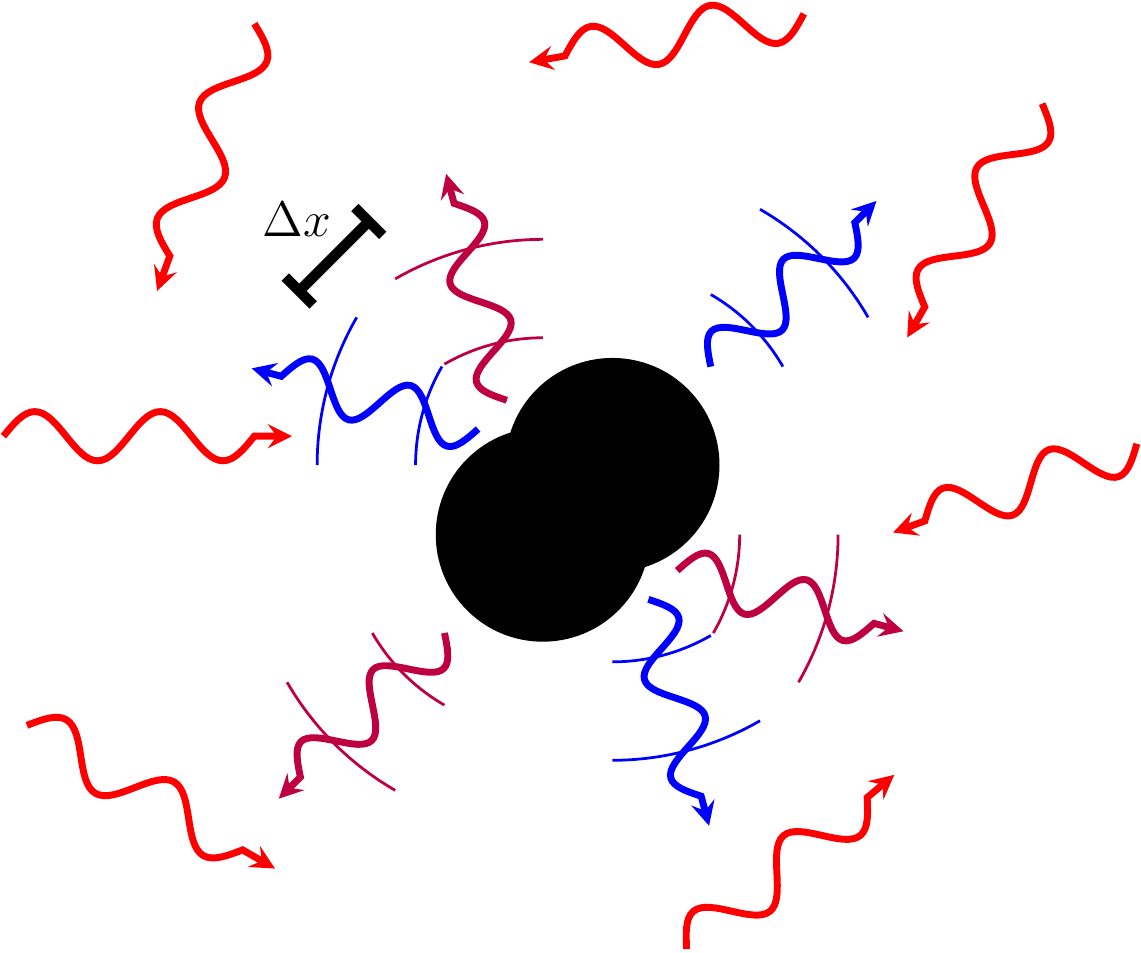}
   \caption{\label{fig:cartoon}A black hole in a superposition of two position states with separation $\Delta x$, immersed in a thermal bath of massless quanta (red). Information about the location of the black hole is carried off by scattered and emitted quanta, shown in blue and purple for the two different positions. }
  \end{center}
\end{figure}

We first consider the case of a Schwarzschild black hole in thermal equilibrium with a radiation bath. Decoherence is caused both by the quanta emitted by the black hole and by the quanta in the external heat bath that are scattered by it (see Fig. \ref{fig:cartoon}). 

The cross section for emission and absorption of Schwarzschild black hole approaches $27 \pi R_s^2$ for high energy quanta of massless species (i.e. the geometrical optics limit) \cite{Page}. When the wavelength of these quanta becomes comparable to the size of the hole, their behavior becomes more complicated and species-dependent. This is because quanta have to penetrate the potential barrier at $\sim 3R_s$ which gives rise to the so called graybody factors for the modes. For the sake of simplicity, we choose to work in the geometrical optics approximation.

In thermal equilibrium the black hole will radiate back as much as falls into the horizon, making the absorption and emission essentially a scattering event from the point of view of decoherence. We therefore adapt the decoherence formalism for photon scattering from dielectric spheres to our current case. The decoherence rate of a dielectric sphere that starts in a superposition of two locations separated by $\Delta x$ and is immersed in a radiation heat bath at temperature $T$ in the dipole approximation ($\Delta x \ll \lambda$, where $\lambda$ is the dominant wavelength of the radiation) is given by \citep{Schlosshauer,JZ,GF}:
\begin{equation}\label{eq:scatter_rate}
\tau_D^{-1}(\Delta x)=\left(16\frac{8! \zeta(9)}{9 \pi} \right)\frac{\tilde{a}^6\Delta x^2(k_B T)^9}{c^8 \hbar^9}
\end{equation}
Above $\tilde a$ is the effective radius of the sphere.
Adapting this for a Schwarzschild black hole in radiation bath at temperature $T=T_H$, we set $\tilde a^2\simeq 27 R_s^2$. The resulting decoherence rate is
then:
\begin{eqnarray}\label{eq:bh_scatter_rate}
\tau_D^{-1}(\Delta x)&=&\left(16\frac{8! \zeta(9)}{9 \pi} \right)\frac{27^3}{(4\pi)^9}\left(\frac{\Delta x}{R_s} \right)^2\left(\frac{c}{R_s} \right)\nonumber \\ &=&d\left(\frac{\Delta x}{R_s} \right)^2\left(\frac{c}{R_s} \right)
\end{eqnarray}
where $d\simeq0.0576$. 
The total rate will be proportional to the number of such species,
and will have to be suitably modified for massive quanta. The basic timescale here is set by the ``light-crossing time'' $R_s/c$.

Neither the decoherence rate nor the corresponding decoherence time:
\begin{eqnarray}\label{eq:bh_scatter_time}
\tau_D(\Delta x)&=&d^{-1}\left(\frac{R_s}{\Delta x} \right)^2\left(\frac{R_s}{c} \right) \nonumber \\ & \simeq& 17.37 \left(\frac{R_s}{\Delta x} \right)^2\left(\frac{R_s}{c} \right)
\end{eqnarray}
depend on the Planck constant $\hbar$. This is unusual as other decoherence rates and times generally depend on $\hbar$. 

Two extreme cases (that mark the two likely limits of the range of applicability of Eqs. (\ref{eq:bh_scatter_rate},\ref{eq:bh_scatter_time})) are worth noting. First, when the size of the Schr{\"o}dinger cat superposition $\Delta x$ is comparable to the black hole radius, $\tau_D$ is of the order of a few black hole light crossing times. Consequently, superpositions of black hole would decohere more slowly when separated by the same distance, or even by the same fraction of their Schwarzschild radius. This may seem surprising (usually larger systems decohere faster), but the temperature of the radiation responsible for decoherence decreases with black hole size, and this effect dominates. For superpositions $\Delta x> R_s$, the dipole approximation (which assumes that the dominant wavelength responsible for decoherence is larger than $\Delta x$) breaks down, and the decoherence rate saturates \cite{GF} (i.e., it does not increase with larger separations, as
Eqs. (\ref{eq:scatter_rate}--\ref{eq:bh_scatter_rate}) would suggest).

The other interesting case where Eqs. (\ref{eq:scatter_rate}--\ref{eq:bh_scatter_time}) will likely break down is when $\Delta x$ becomes equal to Planck length $\ell_p$. In that case the decoherence time is
given by:
\begin{eqnarray}
\tau_D(\ell_p)&=&d^{-1}\left(\frac{R_s}{\ell_p} \right)^2\left(\frac{R_s}{c} \right)=\frac{8}{d}\frac{G^2 M^3}{\hbar c^4}\nonumber \\* &\simeq& 139\frac{G^2 M^3}{\hbar c^4}
\end{eqnarray}
Thus -- assuming the black hole stood still -- it would be localized to Planck length on a timescale comparable to but somewhat shorter than its lifetime due to evaporation into vacuum:
\begin{equation}\label{eq:evap_time}
t_{BH}=5120\pi \frac{G^2 M^3}{\hbar c^4}
\end{equation}
Both expressions assume a single massless mode, and would change accordingly otherwise.

\section*{\label{sec:vacuum}Decoherence in Vacuum}

Next, let us consider decoherence in the case of emission into a vacuum. Aside from changing from effective scattering to emission, this case differs because the black hole will be evaporating and thus have a time dependent temperature. However, assuming the black hole is sufficiently large, we can follow the standard quasi-static formalism\cite{Hawking,Page} and say that the black hole will evaporate slowly compared to the decoherence rate. We will also once again use the geometrical optics limit to simplify our calculation. 

\begin{figure}
 \includegraphics[width=.48\textwidth]{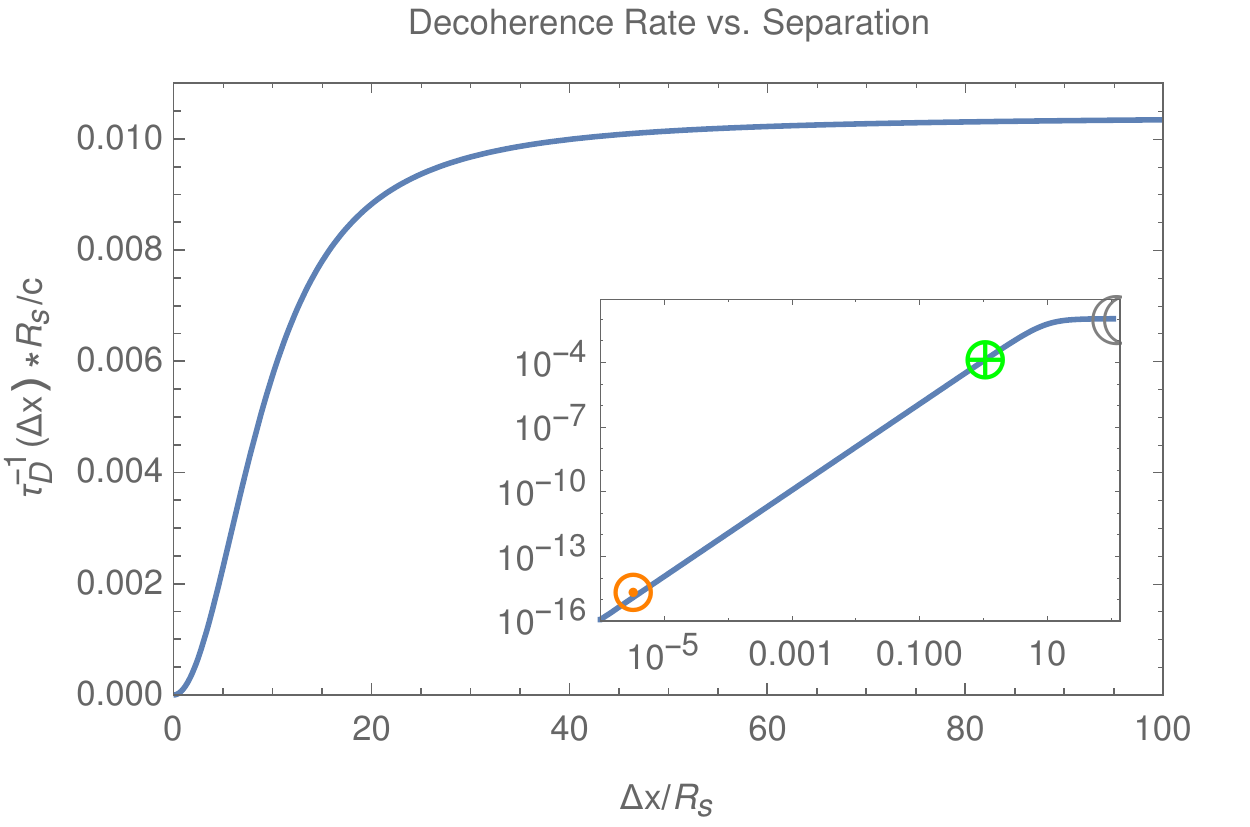}
 \caption{\label{fig:deco}The decoherence rate as a function of $\Delta x/R_s$. Note that this levels off for larger seperations. Inset is the same function plotted on a log-log scale and showing the decoherence rates that correspond to black hole superpositions with separations of $1 \text{cm}$ if the black hole had the mass of the sun($\Sun$), the earth ($\Earth$), and the moon ($\Moon$). The mass, Hawking temperature, and decoherence time for these cases are given below. For the sun $M_\Sun=1.99*10^{30} kg$, $T_{H\Sun}=6.17*10^{-8}K$, and $\tau_{D\Sun}=7.52*10^9 s$. For the earth $M_\Earth=5.97*10^{24} kg$, $T_{H\Earth}=.0205K$, and $\tau_{D\Earth}=2.07*10^{-7} s$. For the moon $M_\Moon=7.35*10^{22} kg$, $T_{H\Moon}=1.67K$, and $\tau_{D\Moon}=1.09*10^{-11} s$.}
\end{figure}
Under these approximations, for the emission of a single massless species we get a decoherence rate of (see the Appendix for details):
\begin{eqnarray}
\label{eq:cold_full}
\tau_D^{-1}&(&\Delta x)=\frac{27 c \zeta (3)}{8 \pi ^4 Rs} \nonumber \\ &+&\frac{27 i c \left(\psi
   ^{(1)}\left(1-\frac{i \Delta x }{4 \pi  R_s}\right)-\psi
   ^{(1)}\left(\frac{i \Delta x }{4 \pi 
   R_s}+1\right)\right)}{8 \pi ^3 \Delta x }.
\end{eqnarray}
Here $\psi^{(1)}\left(y\right)$ is the $n=1$ polygamma function. (See Fig. \ref{fig:deco} for a plot of this function and some test cases.) Note that again, $\hbar$ does not appear anywhere in this expression.

As with the thermal bath case, we will take a moment here to discuss the limiting conditions on this expression. Beginning with the large $\Delta x$ limit, we can explicitly find the asymptotic saturation discussed in the equilibrium case:
\begin{equation}\label{eq:large_cold_limit}
\lim_{\Delta x \to \infty} \tau_D^{-1}(\Delta x)=\frac{27 c \zeta (3)}{32 \pi ^4 Rs}=\Lambda_{total} 
\end{equation}
where $\Lambda_{total}$ is the total emission rate for the massless species (see the Appendix.) In other words, for sufficiently large separations the decoherence time becomes the time to emit a particle. 

In the limit that $\Delta x$ is small compared to $R_s$, Eqn. (\ref{eq:cold_full}) is approximately:
\begin{eqnarray}\label{eq:small_cold_limit}
\tau_D^{-1}(\Delta x \ll R_s) &\simeq& \frac{27\zeta(5)}{256 \pi^6}\left(\frac{\Delta x}{R_s} \right)^2\left(\frac{c}{R_s} \right)\nonumber \\ & \simeq& 1.138*10^{-4}\left(\frac{\Delta x}{R_s} \right)^2\left(\frac{c}{R_s} \right).
\end{eqnarray}
We note that, except for a different numerical prefactor, this decoherence rate has the same form as our approximate result for the equilibrium case. Once again, we expect that this result would break down once $\Delta x$ approaches $\ell_p$, which would give us a decoherence time of:
\begin{eqnarray}
\tau_D(\ell_p)&=&\frac{8 *256\pi^6}{27\zeta(5)}\frac{G^2 M^3}{\hbar c^4} \simeq 22400\pi\frac{G^2 M^3}{\hbar c^4}.
\end{eqnarray}
Note that this is longer than the black hole evaporation time of Eqn. (\ref{eq:evap_time}).

\section*{\label{sec:discussion}Discussion and Conclusions}

The independence of the decoherence rate on $\hbar$ is not the result of either the  dipole or geometrical optics approximations we used. In particular, the geometrical optics approximation that ignores the details of the graybody factors plays no role for the black hole in the heat bath (as Kirchhoff's Law\citep{Kirchhoff} indicates that the combination of scattered and emitted quanta induces as much decoherence as if it were indeed a black body emitter.)

In the case of emission into vacuum, the graybody factors will obviously modify the emitted spectrum, endowing the black hole with color. 
However, these graybody factors depend on the ratio of the wavelength to the Schwarzschild radius in a way that does not introduce any dependence on $\hbar$.

This independence of $\hbar$ contrasts with the ``standard lore'' \citep{Schlosshauer,JZ,GF}. For instance, in quantum Brownian motion the decoherence rate is proportional to $\gamma\left(\Delta x / \lambda_{dB}(T) \right)^2$ where $\gamma$ is the rate of energy loss and $\lambda_{dB}(T)=\hbar\sqrt{2 \pi/(m k_B T)}$ is the thermal de Broglie wavelength. Decoherence is therefore generally faster when $\hbar$ is small compared to $\Delta x \sqrt{m k_B T}$. 

Generalizing to a black hole with angular momentum and charge (the Kerr-Newman case) also does not introduce any $\hbar$ dependence for the decoherence rate for massless, uncharged quanta. Aside from breaking spherical symmetry and adjusting the black hole's radius, this generalization would modify the Bose-Einstein statistics factor for the emissions by adding dependence on the angular mode number $m$ and charge $e$ of the emitted quanta: $\omega \to\omega -m \Omega - e \Phi/\hbar$, where $\Omega$ is the angular velocity of the event horizon and $\Phi$ is the static electric potential from the black hole's charge evaluated at the event horizon. So long as the particles that are being emitted with these statistics have $e=0$, $\hbar$ does not enter into the statistics and thus the resulting decoherence rate remains $\hbar$ independent.

The decoherence rate does, however, depend on $\hbar$ when the black hole is small enough to emit massive particles. This is because they have a lower frequency bound given by $\omega_{min}=\frac{m c^2}{\hbar}$ which is introduced as a cutoff. Additionally, the emission of charged particles from a charged black hole has an additional factor of $\hbar$ in the statistics due to the electromagnetic interaction. 

As our paper addresses decoherence of nonlocal superpositions of black holes, it is natural to enquire whether such superpositions could arise in nature. One plausible candidate would be three-body systems involving one or more black hole. Such many-body systems can evolve chaotically, so that the initial wavepacket of each body would spread exponentially fast at the rate given by Lyapunov exponents\citep{ZurekChaos}.

LIGO (the Laser Interferometer Gravitational-Wave Observatory) has now detected several black hole mergers \cite{Ligo,Ligo_rate}. As black hole ``binaries'' seem to be relatively plentiful, it is possible that there may also black hole triplets with chaotic trajectories that would delocalize the individual black holes on the relevant Lyapunov timescale. Thus, while performing a ``double slit experiment'' with black hole does not seem feasible, dynamics that could lead to nonlocal superpositions may well be present in astrophysical settings. Of course, in such settings other environmental elements will likely play an important role, resulting in decoherence faster than Hawking radiation. Moreover, three-body systems with black holes would exert tidal forces, which may stir up internal degrees of freedom of black holes and accelerate decoherence beyond our estimates. 

We close by noting that the lack of dependence on $\hbar$ of the localization rate--the decoherence rate of spatial superpositions--of black holes may be a consequence of the fact that a black hole is not just an object \emph{in space}, but that it \emph{is} a curved space. One is therefore tempted to speculate that black holes may not need decoherence to be localized.  

\section*{Acknowledgments}
This research was supported by the DoE under the LDRD program at the Los Alamos National Laboratory. Both AA and AA were supported in part by UC Davis.
\section*{Appendix}
\subsection{Decoherence by radiation}

Consider the decoherence of spatial superpositions by radiation emitted into a state $|\chi\rangle$. With the assumption that the state of the emitter is approximately constant, the off diagonal elements of the reduced density matrix of the emitter's position state get suppressed as:
\begin{equation}
\rho(\bm x, \bm x')'=\rho(\bm x, \bm x')\langle \chi (\bm x')|\chi(\bm x)\rangle.
\end{equation}
The change in the reduced density matrix is then:
\begin{equation}
\rho(\bm x, \bm x')'-\rho(\bm x, \bm x')=-\rho(\bm x, \bm x')(1-\langle \chi (\bm x')|\chi(\bm x)\rangle).
\end{equation}
If such an emission happens at some constant rate $\Lambda$, we expect the information about this event to pass observers off at infinity at the same rate. Thus, after some small time $\Delta t$ has passed, we should have a change in the reduced density matrix of:
\begin{eqnarray}
\rho(\bm x, \bm x',\Delta t)&-&\rho(\bm x, \bm x',0)\nonumber \\* &=&-\rho(\bm x, \bm x')\Lambda \Delta t(1-\langle \chi (\bm x')|\chi(\bm x)\rangle).
\end{eqnarray}
Dividing by $\Delta t$ and taking the limit $\Delta t \to 0$, we have:
\begin{equation}
\frac{\partial{\rho(\bm x, \bm x')}}{\partial{t}}=-\Lambda(1-\langle \chi (\bm x')|\chi(\bm x)\rangle)\rho(\bm x, \bm x').
\end{equation}
Our decoherence rate is therefore given by:
\begin{equation}
\tau_D^{-1} = \Lambda(1-\langle \chi (\bm x')|\chi(\bm x)\rangle).
\end{equation}

For Hawking radiation from a large black hole, we generally expect emission of massless species such as photons and gravitons. (Smaller black holes would also radiate massive particles, but we will not treat this case.) These emissions will be into a distribution of states with differing frequencies (or momenta) and angular momentum  for each species. 
 $\Lambda$ will then be the total emission rates for all momenta and mode types for a given species.

\subsection*{Calculating the overlap}
Let us now evaluate the inner product term. Following \citet{hb}, we can express the density matrix for the radiated particles (assuming isotropic emission) in momentum space as:
\begin{equation}
\rho_{radiation}=\int \text{ d} \bm q \frac{p(q)}{4 \pi q^2} | \bm q\rangle \langle \bm q|.
\end{equation}
where $p(q)$ is the probability of emitting a particle with this magnitude of momentum. 

In order to account for the emitter being displaced from the origin, we simply act with  translation operators:  
\begin{equation}
\rho_{radiation}=\int \text{ d} \bm q \frac{p(q)}{4 \pi q^2} e^{-i\hat{\bm q} \cdot \bm x/\hbar}| \bm q\rangle \langle \bm q|e^{i\hat{\bm q} \cdot \bm x'/\hbar}.
\end{equation}

If we define our coordinate system so that $\bm x -\bm x'= \Delta x \hat z$ and then change the integral to spherical coordinates, the trace over this density matrix (the overlap between the states) of emitted particles with different emitter positions is then:
\begin{equation}
\langle\chi(\bm x ')|\chi(\bm x)\rangle= \int \text{ d} q \int \text{ d}\phi \text{ d}\cos(\theta) e^{-i q \Delta x \cos(\theta)/\hbar} \frac{p(q)}{4 \pi} .
\end{equation}
Integrating over the angular variables gives us:
\begin{equation}
\langle\chi(\bm x ')|\chi(\bm x)\rangle= \int\text{ d} q \ \text{}   \text{sinc} \left(\frac{\Delta x p}{\hbar}\right)p(q).
\end{equation} 
In order to match standard notations, let us change variables from momenta $p$ to angular frequency $\omega$:
\begin{equation}
\langle\chi(\bm x ')|\chi(\bm x)\rangle=   \int \text{ d} \omega \text{sinc} \left(\frac{\Delta x \omega}{c}\right) p(\omega).
\end{equation} 
where $p(\omega)=\frac{c}{\hbar}p(q)$. Again following \citet{hb}, we define the rate of particle emission per $\omega$ as:
\begin{equation}
\Lambda(\omega)=  p(\omega) * \Lambda_{total},
\end{equation}
with
\begin{equation}
\Lambda_{total}=\int \text{ d}\omega \Lambda(\omega).
\end{equation}
In this notation, we can express our overlap as:
\begin{equation}
\langle\chi(\bm x ')|\chi(\bm x)\rangle= \frac{1}{\Lambda_{total}} \int \text{ d} \omega \Lambda(\omega)  \text{sinc} \left(\frac{\Delta x \omega}{c}\right).
\end{equation}

\subsection*{Black hole decoherence rate}

Specialising to a Schwarzschild black hole, \citet{Page} tells us that   emission rate per freqency of the Hawking Radiation should be:
\begin{equation}\label{eq:emission_rate}
\Lambda(\omega)=\sum_{s,l,m} \frac{1}{2\pi}\Gamma_{s,l,m}(e^{4\pi \omega R_s/c}-1)^{-1},
\end{equation}
where the $\Gamma$'s are the greybody factors. Using the geometrical optics approximation for a massless species with two polarization degrees of freedom, we have \citep{Page}:
\begin{equation}
\sum_{s,l,m} \Gamma_{s,l,m}=2*\frac{27 R_s^2 \omega ^2}{c^2}.
\end{equation}
Our emission rate per frequency interval (summed over the modes) is then:
\begin{equation}
\Lambda(\omega)=\frac{27 R_s^2 \omega ^2}{\pi c^2}(e^{4\pi \omega {R_s}/c}-1)^{-1}.
\end{equation}
This gives a total emission rate of:
\begin{equation}
\Lambda_{total}=\frac{27 \zeta(3) c}{32 \pi ^4 R_s}.
\end{equation} 
The corresponding overlap is: 
\begin{eqnarray}
&\langle&\chi(\bm x ')|\chi(\bm x)\rangle\nonumber \\* &=& \frac{i \pi  R_s \left(\psi ^{(1)}\left(\frac{i \Delta x }{4 \pi 
   R_s}+1\right)-\psi ^{(1)}\left(1-\frac{i \Delta x }{4
   \pi  R_s}\right)\right)}{\Delta x  \zeta (3)}.
\end{eqnarray}
This gives the decoherence rate of Eqn. \ref{eq:cold_full}.

\bibliography{BH_decoarxiv01}{}

\begin{thebibliography}{17}
\providecommand{\natexlab}[1]{#1}
\providecommand{\url}[1]{\texttt{#1}}
\expandafter\ifx\csname urlstyle\endcsname\relax
  \providecommand{\doi}[1]{doi: #1}\else
  \providecommand{\doi}{doi: \begingroup \urlstyle{rm}\Url}\fi

\bibitem[Hawking(1975)]{Hawking}
S.~W. Hawking.
\newblock Particle creation by black holes.
\newblock \emph{Communications in Mathematical Physics}, 43\penalty0
  (3):\penalty0 199--220, Aug 1975.
\newblock ISSN 1432-0916.
\newblock \doi{10.1007/BF02345020}.
\newblock URL \url{https://doi.org/10.1007/BF02345020}.

\bibitem[Bekenstein(1973)]{Bekenstein}
J.~D. Bekenstein.
\newblock Black holes and entropy.
\newblock \emph{Phys. Rev. D}, 7:\penalty0 2333--2346, Apr 1973.
\newblock \doi{10.1103/PhysRevD.7.2333}.
\newblock URL \url{https://link.aps.org/doi/10.1103/PhysRevD.7.2333}.

\bibitem[Joos et~al.(2003)Joos, Zeh, Kiefer, Giulini, Kupsch, and
  Stamatescu]{Joos_book}
E.~Joos, H.~D. Zeh, C.~Kiefer, D.~Giulini, J.~Kupsch, and I.O. Stamatescu.
\newblock \emph{Decoherence and the Appearance of a Classical World in Quantum
  Theory}.
\newblock Springer Berlin Heidelberg, 2003.
\newblock \doi{10.1007/978-3-662-05328-7}.
\newblock URL \url{https://doi.org/10.1007/978-3-662-05328-7}.

\bibitem[Zurek(2003)]{Zurek_review}
W.~H. Zurek.
\newblock Decoherence, einselection, and the quantum origins of the classical.
\newblock \emph{Rev. Mod. Phys.}, 75:\penalty0 715--775, May 2003.
\newblock \doi{10.1103/RevModPhys.75.715}.
\newblock URL \url{https://link.aps.org/doi/10.1103/RevModPhys.75.715}.

\bibitem[Schlosshauer(2007)]{Schlosshauer}
M.~Schlosshauer.
\newblock \emph{Decoherence and the quantum-to-classical transition}.
\newblock Springer Berlin Heidelberg, 2007.
\newblock \doi{10.1007/978-3-540-35775-9}.
\newblock URL \url{https://doi.org/10.1007/978-3-540-35775-9}.

\bibitem[Zurek(1981)]{Zurek_Pointer}
W.~H. Zurek.
\newblock Pointer basis of quantum apparatus: Into what mixture does the wave
  packet collapse?
\newblock \emph{Phys. Rev. D}, 24:\penalty0 1516--1525, Sep 1981.
\newblock \doi{10.1103/PhysRevD.24.1516}.
\newblock URL \url{https://link.aps.org/doi/10.1103/PhysRevD.24.1516}.

\bibitem[Zurek(1982)]{Zurek_Einselection}
W.~H. Zurek.
\newblock Environment-induced superselection rules.
\newblock \emph{Phys. Rev. D}, 26:\penalty0 1862--1880, Oct 1982.
\newblock \doi{10.1103/PhysRevD.26.1862}.
\newblock URL \url{https://link.aps.org/doi/10.1103/PhysRevD.26.1862}.

\bibitem[Zurek(2014)]{Zurek_Darwin}
W.~H. Zurek.
\newblock Quantum darwinism, classical reality, and the randomness of quantum
  jumps.
\newblock \emph{Physics Today}, 67\penalty0 (10):\penalty0 44--50, 2014.
\newblock \doi{10.1063/PT.3.2550}.
\newblock URL \url{http://dx.doi.org/10.1063/PT.3.2550}.

\bibitem[Bell(1990)]{Bell}
J.~Bell.
\newblock Against `measurement'.
\newblock \emph{Physics World}, 3\penalty0 (8):\penalty0 33, 1990.
\newblock URL \url{http://stacks.iop.org/2058-7058/3/i=8/a=26}.

\bibitem[Page(1976)]{Page}
D.~N. Page.
\newblock Particle emission rates from a black hole: Massless particles from an
  uncharged, nonrotating hole.
\newblock \emph{Phys. Rev. D}, 13:\penalty0 198--206, Jan 1976.
\newblock \doi{10.1103/PhysRevD.13.198}.
\newblock URL \url{https://link.aps.org/doi/10.1103/PhysRevD.13.198}.

\bibitem[Joos and Zeh(1985)]{JZ}
E.~Joos and H.~D. Zeh.
\newblock The emergence of classical properties through interaction with the
  environment.
\newblock \emph{Zeitschrift f{\"u}r Physik B Condensed Matter}, 59\penalty0
  (2):\penalty0 223--243, Jun 1985.
\newblock ISSN 1431-584X.
\newblock \doi{10.1007/BF01725541}.
\newblock URL \url{https://doi.org/10.1007/BF01725541}.

\bibitem[Gallis and Fleming(1990)]{GF}
M.~R. Gallis and G.~N. Fleming.
\newblock Environmental and spontaneous localization.
\newblock \emph{Phys. Rev. A}, 42:\penalty0 38--48, Jul 1990.
\newblock \doi{10.1103/PhysRevA.42.38}.
\newblock URL \url{https://link.aps.org/doi/10.1103/PhysRevA.42.38}.

\bibitem[Kirchhoff(1860)]{Kirchhoff}
G.~Kirchhoff.
\newblock I. {On} the relation between the radiating and absorbing powers of
  different bodies for light and heat.
\newblock \emph{Philosophical Magazine}, 20\penalty0 (130):\penalty0 1--21,
  1860.
\newblock \doi{10.1080/14786446008642901}.
\newblock URL
  \url{http://www.tandfonline.com/doi/abs/10.1080/14786446008642901}.

\bibitem[Zurek(1998)]{ZurekChaos}
Wojciech~H. Zurek.
\newblock Decoherence, chaos, quantum-classical correspondence, and the
  algorithmic arrow of time.
\newblock \emph{Physica Scripta}, T76\penalty0 (1):\penalty0 186, 1998.
\newblock \doi{10.1238/physica.topical.076a00186}.
\newblock URL \url{https://doi.org/10.1238/physica.topical.076a00186}.

\bibitem[et~al.(2016{\natexlab{a}})]{Ligo}
B.{\hspace{0.167em}}P.~Abbott et~al.
\newblock Observation of gravitational waves from a binary black hole merger.
\newblock \emph{Physical Review Letters}, 116\penalty0 (6), feb
  2016{\natexlab{a}}.
\newblock \doi{10.1103/physrevlett.116.061102}.
\newblock URL \url{https://doi.org/10.1103/physrevlett.116.061102}.

\bibitem[et~al.(2016{\natexlab{b}})]{Ligo_rate}
B.~P.~Abbott et~al.
\newblock The rate of binary black hole mergers inferred from advanced ligo
  observations surrounding {GW}150914.
\newblock \emph{The Astrophysical Journal}, 833\penalty0 (1):\penalty0 L1, Nov
  2016{\natexlab{b}}.
\newblock \doi{10.3847/2041-8205/833/1/l1}.
\newblock URL \url{https://doi.org/10.3847/2041-8205/833/1/l1}.

\bibitem[Hornberger et~al.(2004)Hornberger, Sipe, and Arndt]{hb}
K.~Hornberger, J.~Sipe, and M.~Arndt.
\newblock Theory of decoherence in a matter wave talbot-lau interferometer.
\newblock \emph{Phys. Rev. A}, 70:\penalty0 053608, Nov 2004.
\newblock \doi{10.1103/PhysRevA.70.053608}.
\newblock URL \url{https://link.aps.org/doi/10.1103/PhysRevA.70.053608}.

\end{thebibliography}
\bibliographystyle{unsrtnat}

\end{document}